\documentclass[aps,preprint]{revtex4}%
\usepackage{amsfonts}
\usepackage{amsmath}
\usepackage{amssymb}
\usepackage{graphicx}%
\begin{document}
\title[Scaling Symmetries of Radiation Scatterers]{Scaling Symmetries of Scatterers of Classical Zero-Point Radiation}
\author{Timothy H. Boyer}
\affiliation{Department of Physics, City College of the City University of New York, New
York, New York 10031}

\begin{abstract}
Classical radiation equilibrium (the blackbody problem) is investigated by the
use of an analogy. \ Scaling symmetries are noted for systems of classical
charged particles moving in circular orbits in central potentials
$V(r)=-k/r^{n}$ when the particles are held in uniform circular motion against
radiative collapse by a circularly polarized incident plane wave. \ Only in
the case of a Coulomb potential $n=1$ with fixed charge $e$ is there a unique
scale-invariant spectrum of radiation versus frequency (analogous to
zero-point radiation) obtained from the stable scattering arrangement. \ These
results suggest that non-electromagnetic potentials are not appropriate for
discussions of classical radiation equilibrium.

\end{abstract}
\maketitle

\section{Introduction}

The blackbody problem within classical physics has never been solved.
\ Solution requires accurate calculations determining the spectra of random
classical radiation which are in steady-state equilibrium with various
classical mechanical scattering systems. \ Calculations for a few scattering
potentials have been performed. \ Scattering calculations involving the dipole
approximation for particles of very small charge in a harmonic
potential\cite{B1975} or in some nonlinear potentials have been carried
out.\cite{VV}\cite{Nonlin} \ Although the harmonic potential gives no
condition on the radiation spectrum other than spatial isotropy, the other
potentials suggest that the Rayleigh-Jeans spectrum is the only radiation
spectrum giving equilibrium between radiation and matter in classical physics.
\ Also, a class of potentials has been treated when the scattering particle's
linear momentum takes the relativistic form,\cite{Blanco} and again the
conclusion was that equilibrium required the Rayleigh-Jeans spectrum. \ 

It has been suggested for more than thirty years\cite{Nonlin}\ that
understanding of classical radiation equilibrium may require the use of
relativistic scattering systems, specifically particles in Coulomb potentials.
\ More recently it has been noted that the unique value $e$ of electronic
charge may also play a crucial role in classical radiation
equilibrium.\cite{Connecting} \ In the last two decades, these suggestions
have been been greeted with overwhelming rejection by the referees at the
leading physics journals. \ One referee declared that these suggestions are
merely desperate searches for "loopholes" to avoid the obvious conclusion that
classical physics is associated with the Rayleigh-Jeans spectrum. \ Other
referees have suggested that such ideas are so far from currently accepted
physics that they were unpublishable unless a complete solution of the
classical blackbody problem (and indeed of all atomic physics) was presented. \ 

Nevertheless the fact remains that there has never been a classical scattering
calculation involving the Coulomb potential, which is the only classical
potential which has been extended to a fully relativistic system. \ Now the
Coulomb potential is special not only because of its connections to
relativistic physics but also because of its scaling symmetry properties.
\ Since most physicists seem unable to take seriously the need for
relativistic scattering systems for blackbody radiation equilibrium, perhaps
scaling symmetry (which is easy to analyze) may afford an easier glimpse into
the problems of \ classical radiation equilibrium. \ A scaling analysis of
scattering is what we carry out in this article. \ We carry out precise
calculations for a model scattering situation and note the scale invariance
and universal character of the radiation spectrum associated with scattering
involving the Coulomb potential and only the Coulomb potential. \ We believe
the results indicate the serious nature of the remaining "loophole" in the
classical blackbody problem. \ 

\section{Classical Electromagnetic Zero-Point Radiation}

In this article we carry out a model scattering calculation which represents a
crude analogue to treating zero-temperature thermal radiation, classical
electromagnetic zero-point radiation. \ Now apparently many physicists are
still unaware of the concept of classical zero-point radiation, and indeed
some referees at the leading physics journals regularly reject the possibility
of classical electromagnetic zero-point radiation. \ However, classical
zero-point radiation is an intrinsic part of classical electromagnetism which
enters the theory as the homogeneous boundary condition on Maxwell's
equations.\cite{B1975} \ It is required in order to give a classical
electromagnetic description of the experimentally observed van der Waals
forces (Casimir forces) between macroscopic objects at zero
temperature.\cite{Exp} \ Furthermore, classical electromagnetic zero-point
radiation in connection with electric dipole harmonic oscillator systems
provides a classical description of a number of phenomena which are usually
regarded as having only a quantum description, such as van der Waals
forces,\cite{B1975} diamagnetism,\cite{Dia} and thermal effects of
acceleration through the vacuum.\cite{Tacc} \ \ A reader of this manuscript
who can not conceive of classical electromagnetic zero-point radiation may
find it hard to follow the logic of our model calculation. \ The decision to
restrict our analogy to zero-temperature was made so as to simplify the
analysis as much as possible and thereby hopefully make it transparent to readers.

Classical zero-point radiation is the random classical radiation which is
present at the absolute zero of temperature. \ In order to fit the
experimentally observed Casimir forces (as calculated numerous
times\cite{b1975} under various conditions within classical electrodynamics),
it must take the form
\begin{equation}
E_{\omega}=b_{zp}\omega
\end{equation}
where $E_{\omega}$ is the average energy per normal mode of angular frequency
$\omega,$ and $b_{zp}$ is a constant. \ For numerical agreement with the
observed Casimir forces, the constant $b_{zp}$ must take the value%
\begin{equation}
b_{zp}=5.27\times10^{-28}\text{\ erg-sec}%
\end{equation}
\ which is a familiar number.\cite{bval} \ The spectrum of zero-point
radiation is (up to a multiplicative constant) the unique spectrum of random
classical radiation which is Lorentz invariant\cite{Lor} and the unique
spectrum which is scale invariant\cite{scale} under the $\sigma_{ltE^{-1}}%
$-scaling allowed by electromagnetism.\cite{Sigma} \ \ Since each normal mode
of the electromagnetic field at (angular) frequency $\omega$ takes the form of
a harmonic oscillator\cite{RadJ} at frequency $\omega$ and can be describable
in terms of action-angle variables, it follows from Eq. (1) \ and
$E=J_{\omega}\omega$ that the average value of the action variable $J_{\omega
}$ (associated with each normal mode of frequency $\omega)$ takes the same
constant value $b_{zp}$ independent of $\omega$.

Since classical electromagnetic radiation in an enclosure can not bring itself
to equilibrium, the interaction of radiation with scattering systems
represents the crucial element in determining the equilibrium spectrum.
\ Thermal radiation equilibrium involves steady-state behavior for both the
scatterer and the radiation. \ The motion of the charged particles in matter
and the random radiation must fit together so that, on average, there is no
change in the particle energy and no change in the spectrum of the random
radiation. \ Presumably there is some aspect encoded within the scatterers
which produces the universal character found for thermal radiation. \ In this
article we investigate the scattering of radiation by charged particles in a
crude model and find that the Coulomb potential possesses scaling aspects
which allow an associated universal radiation spectrum. \ 

\section{The Model Calculation}

In our simplified model, we consider a single charged particle of mass $m$ and
charge $e$ undergoing uniform circular motion with speed $v=\omega r$ in a
central potential $V\left(  r\right)  =-k/r^{n},$ the motion satisfying
Newton's second law%
\begin{equation}
m\gamma\frac{v^{2}}{r}=n\frac{k}{r^{n+1}}%
\end{equation}
\ where $\gamma=(1-v^{2}/c^{2})^{-1/2}$. \ The orbital angular momentum of the
particle is given by
\begin{equation}
J=rm\gamma v
\end{equation}
Now if the particle undergoing uniform circular motion has charge $e$, then it
will radiate energy into the electromagnetic field at the rate\cite{RadRate}%
\begin{equation}
P=\frac{2}{3}\frac{e^{2}}{c^{3}}\omega^{4}\gamma^{4}r^{2}%
\end{equation}
In order to balance the energy loss to emitted electromagnetic radiation, we
imagine a circularly polarized plane wave of angular frequency $\omega$
propagating along the axis perpendicular to the orbital motion. \ (We may
picture the particle's motion as constrained by two frictionless plane sheets
so that it remains in a fixed orbital plane despite the magnetic force due to
the plane wave, or else we can consider two counter-propagating plane waves so
as to eliminate the magnetic field in the orbital \ plane.) \ The amplitude
$\mathcal{E}_{0}$ of the electric field of the plane wave is taken as the
smallest possible value such that the incident wave will provide the energy
loss given in Eq. (5). \ Thus we require
\begin{equation}
P=\frac{2}{3}\frac{e^{2}}{c^{3}}\omega^{4}\gamma^{4}r^{2}=e\mathcal{E}_{0}v
\end{equation}
or, since $r=\omega/v,$
\begin{equation}
\mathcal{E}_{0}=\frac{2}{3}\frac{e}{c^{3}}\omega^{2}v\gamma^{4}%
\end{equation}
\ with an associated radiation energy density $u$ for the incident plane wave%
\begin{equation}
u=\frac{1}{8\pi}\left(  \mathcal{E}_{0}^{2}+\mathcal{B}_{0}^{2}\right)
=\frac{1}{9\pi}\frac{e^{2}}{c^{6}}\omega^{4}v^{2}\gamma^{8}%
\end{equation}
In this fashion we make an association between a particle motion and and an
electromagnetic field corresponding to a steady-state coherent (in contrast to
random) radiation scattering situation. \ This association can be regarded as
an analogue of the association between the average particle motion and the
electromagnetic spectrum which holds in classical thermal radiation
equilibrium. \ Since both scattering situations involve the known classical
electromagnetic interaction between charged particles and electromagnetic
waves, we expect that exploring one situation may give us insights into the
other. \ 

\section{$\sigma_{ltE^{-1}}$-Scaling Symmetry for Classical Electromagnetism}

We say that a system is $\sigma$-scale invariant or has $\sigma$-scaling
symmetry if the system is mapped onto itself under a dilatation which
multiplies appropriate quantities in the system by a factor of $\sigma.$
\ Although the set of all classical mechanical systems allows independent
scalings of length $\sigma_{l},$ time $\sigma_{t},$ and energy $\sigma_{E},$
classical electromagnetism allows only one $\sigma_{ltE^{-1}}$-scaling
symmetry which connects together the scalings of length, time and
energy.\cite{Connecting} \ The scaling coupling of length and time is required
by the appearance of a fundamental velocity $c$ for electromagnetic radiation,
and the coupling of energy and length is required by the appearance of a
smallest elementary charge $e$ in nature. \ Under $\sigma_{ltE^{-1}}$-scaling,
lengths are scaled as $l\rightarrow l^{\prime}=$ $\sigma_{ltE^{-1}}l,$ times
are scaled as $t\rightarrow t^{\prime}=\sigma_{ltE^{-1}}t,$ and energies are
scaled as $E\rightarrow E^{\prime}=E/\sigma_{ltE^{-1}}.$ \ It follows that all
speeds are $\sigma_{ltE^{-1}}$-scale invariant since $v=l/t\rightarrow
v^{\prime}=(\sigma_{ltE^{-1}}l)/(\sigma_{ltE^{-1}}t)=l/t=v.$ \ Electric charge
$q$ is $\sigma_{ltE^{-1}}$-scale invariant since the potential energy $E$
between two point charges $q$ a distance $r$ apart behaves as $q^{2}%
=Er\rightarrow q^{\prime2}=E^{\prime}r^{\prime}=(E/\sigma_{ltE^{-1}}%
)(\sigma_{ltE^{-1}}r)=q^{2}.$ \ Orbital angular momentum $J$ is $\sigma
_{ltE^{-1}}$-scale invariant since $J=rm\gamma v\rightarrow J^{\prime
}=r^{\prime}m^{\prime}\gamma^{\prime}v^{\prime}=(\sigma_{ltE^{-1}}%
r)(m/\sigma_{ltE^{-1}})\gamma v=rm\gamma v=J$ . \ Electromagnetic fields
$\mathcal{E}$ and $\mathcal{B}$ scale under $\sigma_{ltE^{-1}}$-scaling as
$\mathcal{E}=e/r^{2}\rightarrow\mathcal{E}^{\prime}=e/r^{\prime2}%
=e/(\sigma_{ltE^{-1}}r)^{2}=\mathcal{E}/\sigma_{ltE^{-1}}^{2}.$ \ Mass $m$ is
related to energy $mc^{2}$ and so scales as an energy $m\rightarrow m^{\prime
}=m/\sigma_{ltE^{-1}}.$

\section{$\sigma_{ltE^{-1}}$-Scaling Symmetries for the Model Scattering
Systems}

We are now in a position to note the scaling symmetries for the model
scattering systems described above. \ Under $\sigma_{ltE^{-1}}$-scaling
symmetry, a potential energy function $V(r)=-k/r^{n}$ is mapped to a new
potential energy function $V(r)\rightarrow$ $V^{\prime}(r^{\prime
})=V(r)/\sigma_{ltE^{-1}}=-k^{\prime}/(r^{\prime})^{n}=-k^{\prime}%
/(\sigma_{ltE^{-1}}r)^{n}=-\sigma_{ltE^{-1}}^{-n}k^{\prime}/r^{n}%
=-(\sigma_{ltE^{-1}}^{-n}k^{\prime}/k)k/r^{n},$ or $V(r)=-(\sigma_{ltE^{-1}%
}^{1-n}k^{\prime}/k)k/r^{n}.$ Form invariance requires that
\begin{equation}
k^{\prime}=\sigma_{ltE^{-1}}^{n-1}k
\end{equation}
\ Thus the potential energy function retains its form under a $\sigma
_{ltE^{-1}}$-scaling transformation if and only if the constant $k$ appearing
in the potential energy is also transformed to $k^{\prime}=\sigma_{ltE^{-1}%
}^{n-1}k.$ \ Thus there is exactly one potential energy function which is
$\sigma_{ltE^{-1}}$-scale invariant $k^{\prime}=k$, and that is the potential
energy function where $n=1,$ namely the Coulomb potential $V(r)=-k/r.$ \ If we
write the potential function in terms of elementary charges of magnitude $e$,
this is $V(r)=-e^{2}/r.$

The $\sigma_{ltE^{-1}}$-scale invariance associated with the Coulomb potential
reappears in the equation for the orbital speed of the particle. \ Thus
combining equations (3) and (4) for a general potential $V(r)=-k/r^{n}$, we
can eliminate the orbital radius $r$ so as to obtain an equation connecting
the orbital speed $v$ and the orbital angular momentum $J$ in terms of the
particle mass $m$ and the potential parameter $k$%
\begin{equation}
v^{2-n}\gamma^{1-n}=\frac{nk}{J^{n}}m^{n-1}%
\end{equation}
The Coulomb potential with $n=1$ is very special because in this case (and
only in this case) the particle mass $m$ disappears from Eq. (10) for the
orbital speed giving%
\begin{equation}
v=\frac{k}{J}=\frac{e^{2}}{J}%
\end{equation}
This last equation is clearly $\sigma_{ltE^{-1}}$-scale invariant since $v$,
$e,$ and $J$ are all invariant.

This matter of $\sigma_{ltE^{-1}}$-scale invariance reappears in the
scattering spectrum given by Eqs. (7) and (8) and gives a universal character
to the spectrum. \ The electric field $\mathcal{E}_{0}$ appearing in Eq. (7)
can be regarded as a function of the angular frequency $\omega$ of the wave
(which exactly matches the angular frequency of the orbital motion of the mass
$m$ in the potential $V(r)$) and of the orbital particle speed $v$. \ For the
Coulomb potential (and only for the Coulomb potential), the orbital speed $v$
as given in Eq. (11) is independent of the particle mass $m.$ $\ $Thus the
equations Eq. (7) and (8) for the electric field and the electromagnetic
energy density become
\begin{equation}
\mathcal{E}_{0}=\frac{2}{3}\frac{e}{c^{3}}\omega^{2}\left(  \frac{e^{2}}%
{J}\right)  \left[  1-\left(  \frac{e^{2}}{Jc}\right)  ^{2}\right]  ^{-2}%
\end{equation}
and
\begin{equation}
u=\frac{1}{8\pi}\left(  \mathcal{E}_{0}^{2}+\mathcal{B}_{0}^{2}\right)
=\frac{1}{9\pi}\frac{e^{2}}{c^{6}}\omega^{4}\left(  \frac{e^{2}}{J}\right)
^{2}\left[  1-\left(  \frac{e^{2}}{Jc}\right)  ^{2}\right]  ^{-4}%
\end{equation}
We notice that the radiation spectrum given by Eqs. (7) and (8) makes no
reference whatsoever to the orbital motion of any charged particle $m$ other
than through the angular momentum $J.$ \ The spectrum associates an electric
field $\mathcal{E}_{0}$ and corresponding electromagnetic energy density $u$
with a given frequency $\omega$ and with fundamental constants $e$ and $c$
provided that the value of any orbital angular momentum $J$ is chosen as a
constant, perhaps chosen as the same constant as appeared above in Eq. (2).
\ Thus we have arrived at a universal spectrum of just the sort which we would
want to associate with thermal radiation at zero temperature, classical
electromagnetic zero-point radiation. \ We also emphasize that the constant
$k=e^{2}$ can not be allowed to change continuously or the universal character
of the spectrum will be lost. \ 

Under $\sigma_{ltE^{-1}}$-scale transformation, the spectrum given in Eqs. (7)
and (8) is invariant. \ The electric field transforms as $\mathcal{E}%
_{0}\rightarrow\mathcal{E}_{0}^{\prime}=\mathcal{E}_{0}/\sigma_{ltE^{-1}}^{2}$
while the energy per unit volume transforms as $u\rightarrow u^{\prime
}=u/\sigma_{ltE^{-1}}^{4};$ these transformation forms are matched by the
powers of $\omega$ appearing on the right-hand sides of the equations while
all the remaining parameters are $\sigma_{ltE^{-1}}$-scale invariant. \ 

Perhaps it is helpful to characterize the uniqueness of the scattering
spectrum (7) in the Coulomb case in a different way. \ If one rescales units
according to $\sigma_{ltE^{-1}}$-scaling, the value of $k$ appearing in the
potential energy function $V(r)=-k/r^{n}$ changes with the units for every
case except $n=1$. \ Thus it is natural to regard $k$ as a parameter available
for an adiabatic change in mechanics, except for the Coulomb potential
function where the parameter $e^{2}=k$ is $\sigma_{ltE^{-1}}$-scale invariant
and can be chosen as fixed. \ The angular momentum $J$ of a particle in orbit
in the potential $V(r)$ is an action variable and so is an adiabatic invariant
which does not change with an adiabatic change of the parameter $k.$ \ Thus
for a general value of $n$, the different values of $k$ for fixed mass $m$
will involve different frequencies $\omega$ and different scattering electric
fields $\mathcal{E}_{0}.$ \ This generates a spectrum connecting electric
field strength $\mathcal{E}_{0}$ and frequency $\omega$ for each fixed value
of mass $m.$ \ However, in general, different values of mass $m$ will generate
different spectral connections between $\mathcal{E}_{0}$ and $\omega$. \ Now
we remember that thermal radiation equilibrium involves a unique equilibrium
spectrum which is independent of the mass $m$ of the charged particle which is
scattering the radiation. \ In our crude analogue to thermal radiation
scattering, we have found that the Coulomb potential energy function with
unique charge $e$ is associated with a unique spectrum connecting
$\mathcal{E}_{0}$ and $\omega$ independent of the mass $m$ of the orbiting
particle. \ Moreover, the Coulomb scattering potential is the only potential
energy function of the form $V(r)=-k/r^{n}$ which gives such a unique
spectrum. \ 

\section{Scaling Symmetries for Scatterers of Classical Zero-Point Radiation}

The $\sigma_{ltE^{-1}}$-scaling symmetries which we have noted above for our
model scattering calculation reappear in the scattering of random radiation at
zero-temperature corresponding to classical electromagnetic zero-point
radiation. \ The zero-point radiation spectrum is $\sigma_{ltE^{-1}}$-scale
invariant. \ Thus the two-field correlation function is given by\cite{Sigma}%
\begin{align}
&  \left\langle F_{zp}^{\mu\nu}(x)F_{zp}^{\sigma\rho}(y)\right\rangle
\nonumber\\
&  =\left(  g^{\mu\sigma}\partial_{x}^{\nu}\partial_{y}^{\rho}-g^{\mu\rho
}\partial_{x}^{\nu}\partial_{y}^{\sigma}-g^{\nu\sigma}\partial_{x}^{\mu
}\partial_{y}^{\rho}+g^{\sigma\rho}\partial_{x}^{\mu}\partial_{y}^{\nu
}\right)  \frac{2b_{zp}c}{\pi(x-y)^{2}}%
\end{align}
where $F_{zp}^{\mu\nu}(x)$ is the electromagnetic field tensor, $g^{\mu\nu}$
is the metric for Minkowski spacetime, $x$ and $y$ are spacetime
displacements, and $b_{zp}$ is the constant appearing in Eq. (2) which has the
units of angular momentum. \ Under a $\sigma_{ltE^{-1}}$-scale transformation,
the correlation function is invariant\cite{Sigma} since the the four inverse
powers of $\sigma_{ltE^{-1}}$associated with the two factors of
electromagnetic field on the left-hand side are matched by the four inverse
powers of space or time on the right-hand side.

The scattering equations for a particle of charge $e$ in a Coulomb potential
function $V(r)=-e^{2}/r$ can be given in manifestly Lorentz-covariant form.
The particle motion follows the relativistic form of Newton's second law
including radiation reaction (which is usually known as the Lorentz-Dirac
equation)\cite{Teit}%
\begin{equation}
m\frac{d^{2}x^{\mu}}{d\tau^{2}}=eF_{Coul}^{\mu\nu}\frac{dx_{\nu}}{d\tau}%
+\frac{2}{3}\frac{e^{2}}{c^{3}}\left(  \frac{d^{3}x^{\mu}}{d\tau^{3}}+\frac
{1}{c^{2}}\frac{dx^{\mu}}{d\tau}\frac{d^{2}x^{\nu}}{d\tau^{2}}\frac
{d^{2}x_{\nu}}{d\tau^{2}}\right)  +eF_{ZP}^{\mu\nu}\frac{dx_{\nu}}{d\tau}%
\end{equation}
where $F_{Coul}^{\mu\nu}$ gives the electromagnetic fields of the Coulomb
potential, the term involving $2e^{2}/(3c^{3})$ gives the radiation damping,
and $F_{zp}^{\mu\nu}$is the random zero-point radiation with the correlation
function given in Eq. (14). \ The scattered electromagnetic fields can be
written in terms of the potentials as $F^{\mu\nu}=\partial^{\mu}A^{\nu
}-\partial^{\nu}A^{\mu},$ where the potentials satisfy the wave equation with
source $J^{\mu}$ given by the scattering charged particle
\begin{equation}
\square^{2}A^{\mu}=\frac{4\pi}{c}J^{\mu}%
\end{equation}
with solution%
\begin{equation}
A^{\mu}(x)=A^{in\mu}(x)+\frac{1}{c^{2}}%
{\displaystyle\int}
d^{4}x^{\prime}D(x-x^{\prime})J^{\mu}%
\end{equation}
where $A^{in\mu}$ is the vector potential for the incoming zero-point
radiation $F_{zp}^{\mu\nu}=\partial^{\mu}A^{in\nu}-\partial^{\nu}A^{in\mu},$
while $D(x-x^{\prime})=2c\theta(x^{0}-x^{\prime0})\delta\lbrack(x-x^{\prime
})^{2}]$ is the retarded Green function for the scalar wave equation, and
$J^{\mu}$ is the current density for the scattering charged particle $J^{\mu
}(x)=ec%
{\displaystyle\int}
d\tau(dx_{e}^{\mu}/d\tau)\delta^{4}[x-x_{e}(\tau)].$

Since under a scaling transformation, the mass $m$ transforms inversely with
$\sigma_{ltE^{-1}},$ while $x^{\mu}$ and $\tau$ transform directly with
$\sigma_{ltE^{-1}},$ it follows that every term in Eq. (15) scales as
$\sigma_{ltE^{-1}}^{-2}$ and hence that the equation is $\sigma_{ltE^{-1}}%
$-scale invariant provided the mass is transformed. \ The wave equation (16)
involves three inverse powers of $\sigma_{ltE^{-1}}$ on each side and hence is
$\sigma_{ltE^{-1}}$-scale invariant. \ Similarly, every term in Eq. (17)
scales as $\sigma_{ltE^{-1}}^{-1}\,\ $and hence that equation is also
$\sigma_{ltE^{-1}}$-scale invariant. \ Since the scattering equations are
scale invariant provided the mass $m$ is transformed, we expect that the
$\sigma_{ltE^{-1}}$-scale invariance of the incoming zero-point radiation will
be preserved for the Coulomb potential, just as the spectrum in our model
calculation is $\sigma_{ltE^{-1}}$-scale invariant independent of particle
mass $m$. \ However, there has been no complete calculation showing the
invariance of zero-point radiation under scattering by a charged particle in a
Coulomb potential.

In the physics literature, all of the scattering calculations for random
classical radiation involve potentials other than the Coulomb potential. \ In
the inertial frame where the radial force on the orbiting particle takes the
form $F=nk/r^{n+1},$ the particle equation of motion corresponding to Eq. (15)
for a general potential function will not be $\sigma_{ltE^{-1}}$-scale
invariant since invariance in form requires that the constant $k$ be
transformed with $1-n$ factors of $\sigma_{ltE^{-1}},$ as in Eq. (9). \ Since
for a general potential function the equations describing the radiation
scattering are not $\sigma_{ltE^{-1}}$-scale invariance, it should come as no
surprise that the $\sigma_{ltE^{-1}}$-scale invariance of the income
zero-point radiation is not preserved, just as it is not preserved in our
model calculation if $n\neq1$. \ Rather, the Rayleigh-Jeans spectrum, which is
not $\sigma_{ltE^{-1}}$-scale invariant, seems to be the spectrum which is
invariant for (at least dipole) scattering by more general classical
potentials.\cite{VV}\cite{Nonlin}\cite{Blanco}

It is also worth noting that zero-point radiation is the unique spectrum of
random radiation which is Lorentz invariant.\cite{Lor} \ Equation (14) gives
the manifestly Lorentz covariant two-point correlation function for classical
electromagnetic zero-point radiation. \ In Eqs. (15)-(17) we have given the
manifestly Lorentz-covariant form for the scattering equations. \ We suggest
again that the Lorentz-invariant character of zero-point radiation is likely
to be preserved by such a scattering system. \ On the other hand, a general
mechanical scatterer breaks the Lorentz-invariance of the scattering
equations. \ It does not seem surprising that the Lorentz-invariant zero-point
spectrum is transformed toward a non-Lorentz-invariant form by scattering
systems which are not Lorentz invariant.

\section{Conclusion}

In this article we have carried out a specific model calculation for the
coherent electromagnetic radiation spectrum of minimum intensity which will
hold a charged particle in a circular orbit in a potential function
$V(r)=-k/r^{n}.$ \ We find that the radiation spectrum is $\sigma_{ltE^{-1}}%
$-scale invariant and independent of the particle mass for a fixed value of
angular momentum only in the case where the Coulomb potential with fixed
charge $e^{2}$ provides the basic mechanical behavior for the orbit. \ We have
noted the $\sigma_{ltE^{-1}}$-scale invariance of the electromagnetic
radiation and the scattering equations for electromagnetic radiation when the
Coulomb potential is used. \ We believe that our arguments strongly suggest
that the Coulomb potential may provide an electromagnetic scatterer which
leaves invariant the zero-point radiation spectrum. \ Rather than being merely
a "loophole" in the problem of classical radiation equilibrium, this
calculation is the fundamental classical calculation which needs to be performed.

However, nonrelativistic quantum physics apparently has such a complete hold
on the minds of many physicists, that they reject the possibility that some
classical calculations still need to be done to clarify the distinctions
between classical and quantum phenomena.


\begin{thebibliography}{99}                                                                                               %


\bibitem {B1975}T. H. Boyer, "Random electrodynamics: The theory of classical
electrodynamics with classical electromagnetic zero-point radiation," Phys.
Rev. D \textbf{11}, 790-808 (1975). \ See the appendix for the stability of
the zero-point radiation under scattering by an electric dipole oscillator.

\bibitem {VV}J. H. Van Vleck, "The Absorption of Radiation by Multiply
Periodic Orbits, and its Relation to the Correspondence Principle and the
Rayleigh-Jeans Law. Part II Calculation of Absorption by Multiply Periodic
Orbits," Phys. Rev. \textbf{24}, 347-365 (1924).

\bibitem {Nonlin}T. H. Boyer, "Equilibrium of random classical electromagnetic
radiation in the presence of a nonrelativistic nonlinear electric dipole
oscillator," Phys. Rev. D \textbf{13}, 2832-2845 (1976). \ T. H. Boyer,
"Statistical equilibrium of nonrelativistic multiply periodic classical
systems and random classical electromagnetic radiation," Phys. Rev. A
\textbf{18}, 1228-1237 (1978).

\bibitem {Blanco}R. Blanco, L. Pesquera, and E. Santos, "Equilibrium between
radiation and matter for classical relativistic multiperiodic systems.
Derivation of Maxwell-Boltzmann distribution from Rayleigh-Jeans spectrum,"
Phys. Rev. D \textbf{27}, 1254-1287 (1983); "Equilibrium between radiation and
matter for classical relativistic multiperiodic systems II. Study of radiative
equilibrium with Rayleigh-Jeans radiation," Phys. Rev. D \textbf{29},
2240-2254 (1984). \ These articles suggest that a relativistic scatterer again
leads to the Rayleigh-Jeans spectrum as the equilibrium spectrum for classical
radiation. \ The calculations involve the use of the relativistic expression
for the linear momentum of the orbiting particle in a general class of
potentials $V(r)$ which excludes the Coulomb potential. \ Such systems are not
relativistic systems. \ Only the Coulomb potential has been extended to fully
relativistic theory. \ This question of obtaining relativistic theories from
mechanical potentials is discussed by T. H. Boyer, "Concerning Potential
Functions in Relativistic and Nonrelativistic Accelerated Coordinate Frames,"
submitted for publication.

\bibitem {Connecting}T. H. Boyer, "Connecting blackbody radiation, relativity,
and discrete charge in classical electrodynamics," Found. Phys. \textbf{37},
999-1026 (2007). (physics/0605003) 

\bibitem {Exp}M. J. Sparnaay, "Measurement of the attractive forces between
flat plates," Physica \textbf{24}, 751-764 (1958). \ S. K. Lamoreaux,
"Demonstration of the Casimir force in the 0.6 to 6 $\mu$m range," Phys. Rev.
Lett. \textbf{78}, 5-8 (1997); \textbf{81}, 5475-5476 (1998). \ U. Mohideen,
"Precision measurement of the Casimir force from 0.1 to 0.9 $\mu$m," Phys.
Rev. Lett. \textbf{81}, 4549-4552 (1998). \ H. B. Chan, V. A. Aksyuk, R. H.
Keiman, and F. Capasso, "Quantum mechanical actuation of
microelectromechanical systems by the Casimir force," Science \textbf{291},
1941-1944 (2001). \ G. Bressi, G. Carugno, R. Onofrio, and G. Ruoso,
"Measurement of the Casimir force between parallel metallic surfaces," Phys.
Rev. Lett. \textbf{88}, 0441804(4) (2002).

\bibitem {Dia}T. H. Boyer, "Diamagnetism of a free particle in classical
electron theory with classical electromagnetic zero-point radiation," Phys.
Rev. A \textbf{21}, 66-72 (1980).

\bibitem {Tacc}T. H. Boyer, "Thermal effects of acceleration through random
classical radiation," Phys. Rev. D \textbf{21}, 2137-2148 (1980); "Thermal
effects of acceleration for a classical dipole oscillator in classical
electromagnetic zero-point radiation," Phys. Rev. D \textbf{29}, 1089-1095 (1984).

\bibitem {b1975}See for example, T. H. Boyer, "Van der Waals Forces and
Zero-Point Energy for Dielectric and Permeable Materials," Phys. Rev. A
\textbf{9}, 2078-2084 (1974) and "Temperature Dependence of Van Der Waals
Forces in Classical Electrodynamics," Phys. Rev. A \textbf{11}, 1650-1663 (1975)

\bibitem {bval}The classical constant $b_{zp}$ clearly takes the same value as
$(1/2)\hbar.$ \ However, $b_{zp}$ enters in purely classical electromagnetic
theory and has nothing to do with ideas of energy quanta.

\bibitem {Lor}T. W. Marshall, "Statistical Electrodynamics," Proc. Camb. Phil.
Soc. \textbf{61}, 537-546 (1965); T. H. Boyer, "Derivation of the Blackbody
Radiation Spectrum without Quantum Assumptions," Phys. Rev. \textbf{182},
1374-1383 (1969).

\bibitem {scale}T. H. Boyer, "Scaling Symmetry and Thermodynamic Equilibrium
for Classical Electromagnetic Radiation," Found. Phys. \textbf{19}, 1371-1383 (1989).

\bibitem {Sigma}T. H. Boyer, "Conformal Symmetry of Classical Electromagnetic
Zero-Point Radiation," Found. Phys. \textbf{19}, 349-365 (1989).

\bibitem {RadJ}See, for example, E. A. Power, \textit{Introductory Quantum
Electrodynamics }(American Elsevier, New York 1964), pp. 18-22.

\bibitem {RadRate}J. D. Jackson, \textit{Classical Electrodynamics, 2nd. ed.,}
(Wiley, New York 1975), p. 665.

\bibitem {Teit}See for example, C. Teitelboim, D. Villarroel, and Ch. G. van
Weert, "Classical Electrodynamics of Retarded Fields and Point Particles,"
Riv. de Nuovo Cimento \textbf{3}, 1-64 (1980).
\end{thebibliography}
\end{document}